\documentstyle[aps,prl,multicol,epsf]{revtex}
\begin{document}
\draft

\title{Optimal orbitals from energy fluctuations in correlated wave functions}

\author{Stephen Fahy and Claudia Filippi}
\address{Department of Physics, University College, Cork, Ireland}
\date{\today}
\maketitle
\begin{abstract}

A quantum Monte Carlo method of determining Jastrow-Slater wave 
functions for which the energy is stationary with respect to variations 
in the single-particle orbitals is presented.          
A potential is determined by a least-squares fitting of fluctuations in 
the energy with a linear combination of one-body operators.
This potential is used in a self-consistent scheme for the orbitals 
whose solution ensures that the energy of the correlated wave function 
is stationary with respect to variations in the orbitals. The method is 
feasible for atoms, molecules, and solids and is demonstrated for the 
carbon and neon atoms.

\end{abstract}

\begin{multicols}{2}
\setcounter{collectmore}{5}

Over the past decade, quantum Monte Carlo (QMC) 
methods~\cite{CA,UWW,FWL1,LCM,CMR,R,FU,W} have been used
to calculate the structural and electronic properties of a variety
of atoms, molecules, clusters and solids.
For systems with large numbers of electrons, QMC methods at present 
provide the most accurate benchmark calculations of structural energies. 
In both variational (VMC) and diffusion (DMC) calculations, a key
step is the construction of a trial correlated 
many-electron wave function.
In many such calculations, the trial wave function $\Psi$ is chosen 
to be of the Jastrow-Slater form~\cite{FWL1,LCM,CMR,R,W,H}, i.e., 
$\Psi = {\cal J} D$, where $D$ is a determinant of single-particle
orbitals and ${\cal J}$ is the (positive) Jastrow correlation factor.
Although considerable progress has been made (principally using the 
variance minimization approach) in the numerical construction of 
optimal Jastrow factors~\cite{UWW,FU,W,H}, relatively 
little attention has been given to the physical understanding and 
numerical optimization of the antisymmetric part of the wave function.
In few-electron systems, variance minimization has been applied to
the optimization of the determinant~\cite{FU} but, in larger 
systems, local density functional (LDA) or Hartree-Fock (HF) orbitals 
have generally been used as the only practical and numerically accurate
way of constructing the determinantal part of the wave function~\cite{FWL1,LCM,CMR,R,W}.
It has not been 
clear why LDA (or HF) orbitals, which have little formal justification
in this context, do so well or how one might in practice do better.
A physical understanding of this issue and a practical method of
approach to the calculation of such orbitals is bound to be 
particularly important to a wide and successful application of QMC 
methods.

In this Letter, a new iterative method is demonstrated which successively
updates the determinant $D$ in Jastrow-Slater wave functions so that, 
at convergence, the energy of the full correlated wave function 
is stationary with respect to variations in the single-particle orbitals
in the Slater determinant.
The method is cast in the framework of a self-consistent field problem
for the determinant and can make use of the standard numerical
codes (either LDA or HF), combined with VMC sampling 
methods of many-body wave functions.

We will first derive Euler-Lagrange equations satisfied when the
energy is stationary with respect to the single-particle orbitals in 
the determinant. 
Then, we will present the numerical approach for an iterative 
scheme which solves those equations exactly~\cite{EuLa}, demonstrate it in some 
numerical examples, and show that it can be conveniently combined 
with existing variance minimization methods for the optimization 
of the Jastrow factor.

\noindent{\it Euler-Lagrange Equations:}
We assume that ${\cal J}$ is held fixed and that only the single-particle 
orbitals $\phi_i$ of the determinant $D$ are varied. 
A general infinitesimal variation of the orbital $\phi_i$ can be written
as $\phi_i \rightarrow \phi_i + \sum_j \eta_{ji} \phi_j$ where
$\eta_{ji}$ are infinitesimal coefficients and
the sum is over a set of orthonormal orbitals which 
excludes all $\phi_j$ already in $D$.
The corresponding variation in $D$ is given by
$D \rightarrow [ 1 + \sum_{i,j} \eta_{ji} c^\dagger_j c_i ] D$
where $c^\dagger_j$ and $c_i$ are the fermion creation and destruction
operators for the orbitals $j$ and $i$, respectively.
The energy is stationary with respect to all variations of the orbitals
$\phi_i$ if and only if the Euler-Lagrange equations, 
\begin{eqnarray}
\Delta E_{ji} \equiv {\partial \over \partial \eta_{ji} }
{\langle \Psi | {\cal H} | \Psi \rangle \over \langle \Psi | \Psi \rangle }
 = 0,
\label{eqn1}
\end{eqnarray}
are satisfied for all $i$ and $j$. ${\cal H}$ is the many-body Hamiltonian.
Explicit evaluation of the derivatives shows that
\begin{eqnarray}
\Delta E_{ji} =
\langle\Psi|({\cal H}-\bar E)\left[{c^\dagger_j c_i D \over D}\right] |\Psi\rangle,
\end{eqnarray}
where $ \bar E = \langle \Psi | {\cal H} | \Psi \rangle $ and $|\Psi\rangle$ 
is assumed normalized.

By considering arbitrary linear combinations of the variations 
$\eta_{ji}$, we can see that solving the set of
Euler-Lagrange equations (\ref{eqn1}) is equivalent to requiring that
\begin{eqnarray}
\Delta E_{\cal O} \equiv 
\langle\Psi|({\cal H}-\bar E ) \left[{{\cal O}D\over D}\right]|\Psi\rangle = 0
\label{eqn3}
\end{eqnarray} 
for all possible one-body operators $\cal O$.
As discussed below, it is sometimes convenient or sufficient to 
consider a restricted class of variations of the orbitals in $D$ so 
that Eq.~(\ref{eqn3}) is satisfied only for a restricted class of one-body 
operators. 

\noindent {\it Iterative Solution of the Euler-Lagrange Equations:}
We wish to find $D$ such that $\Delta E_{{\cal O}_k} = 0$
for a set of $n$ one-body operators $\{{\cal O}_k\}$.
We sample $N_c$ configurations $\{R(i)\}$ with local energies 
$\{E(i)={\cal H}\Psi(R(i))/\Psi(R(i))\}$ from the square of the wave function 
$\Psi={\cal J} D$.
We perform a least-squares fit of the local energies with the sum
$ E_0 + \sum_{k=1}^n V_k {\cal O}_k (i) $, where 
$E_0$ and $V_k$ are fitting parameters and
${\cal O}_k(i) =  {{\cal O}_k D(R(i))/ D(R(i))}$. 
In the limit of $N_c \rightarrow\infty$, 
this is equivalent to minimizing the integral
\begin{eqnarray}
\langle\Psi|\left\{{\cal H}- E_0 - \sum_{k=1}^n V_k {{\cal O}_k D\over D} \right\}^2 
|\Psi \rangle
\label{eqn4}
\end{eqnarray}
with respect to the fitting parameters. This is in turn equivalent to 
solving the set of linear equations:
\begin{eqnarray}
\sum_{k=1}^{n} V_k \, \langle \Delta {\cal O}_k \Delta {\cal O}_l \rangle 
= \langle \Delta E \Delta {\cal O}_l\rangle,\;\;{\rm for}~ l=1,\dots,n,
\label{eqn5}
\end{eqnarray}
where $\langle \cdot \rangle$ denotes the average over the $N_c$ configurations, 
$\Delta {\cal O}_k(i) = {\cal O}_k(i)-\langle {\cal O}_k \rangle$
and  $\Delta E(i) = E(i)-\langle E \rangle$.
For $N_c \rightarrow \infty$, $\langle \Delta E \Delta {\cal O}_l  \rangle  
\rightarrow \Delta E_{{\cal O}_l}$ and
{\it the fitting coefficients $V_k$ are all zero if and
only if all} $\Delta E_{{\cal O}_k} = 0 $.

Suppose we determined a set of $V^{(1)}_k$ for a wave function 
$\Psi^{(1)}={\cal J} D^{(1)}$ according to the above procedure.
How do we use the coefficients $V^{(1)}_k$ to obtain the orbitals 
$\phi^{(2)}_i$ for the determinant $D^{(2)}$ of the next iteration? 

Let's first suppose we have 
a {\it non-interacting} system with Hamiltonian ${\cal H}_{\rm eff}$ and
eigenfunctions $\phi_i$ but that we start from the eigenstates $\phi^{(1)}_i$
of an {\it incorrect} Hamiltonian ${\cal H}^{(1)}={\cal H}_{\rm eff}-
\sum_{k=1}^n A_k {\cal O}_k $.
If we follow the above fitting procedure 
(Eq.~\ref{eqn5}) and determine the coefficients $V_k$, it is easy to
see that $V_k = A_k$ for all $k$ and ${\cal H}_{\rm eff}$ may be found (if not 
known in advance) as ${\cal H}_{\rm eff} =  {\cal H}^{(1)} + \delta {\cal H}$
where $\delta {\cal H} = \sum_{k=1}^n V_k {\cal O}_k$.
The correct single-particle orbitals $\phi_j$ are then eigenstates 
of the Hamiltonian ${\cal H}^{(1)} + \delta {\cal H}$.

Motivated by this argument for non-interacting systems, we construct the 
determinant $D^{(1)}$ of the correlated wave function $\Psi^{(1)}$ from 
orbitals $\phi^{(1)}_i$ which are eigenstates of a non-interacting Hamiltonian 
${\cal H}^{(1)}$. We compute $V^{(1)}_k$ from Eq.~\ref{eqn5} and, as in the 
non-interacting case, use $\delta {\cal H}^{(1)}=\sum_{k=1}^n V^{(1)}_k {\cal O}_k$
as an increment to the physical external potential $V_{ext}$ in ${\cal H}^{(1)}$
to determine a set of orbitals $\phi^{(2)}_i$.
A new increment $\delta {\cal H}^{(2)}$ is similarly derived 
from ${\cal J}D^{(2)}$ and the 
external potential $V_{ext} + \delta {\cal H}^{(1)} + \delta {\cal H}^{(2)}$ is 
used for the next iteration to obtain the orbitals $\phi^{(3)}_i$.
Convergence is reached when $\delta {\cal H}^{(i)}$ is negligible.
For accelerated convergence,
the orbitals $\phi^{(m)}_i$ at the $m$th iteration are determined by
performing a standard self-consistent LDA calculation with the external
potential $V_{ext} + \sum_{l=1}^{m-1} \delta {\cal H}^{(l)}$.
The scheme converges within two or three iterations for the applications 
studied here.
It should be noted that the LDA potential here is used purely for 
computational convenience and that the final orbitals $\phi^{(m)}_i$,
the Hartree potential and the final energy-fluctuation potential (EFP) part 
of the non-interacting Hamiltonian, ${\cal H}^{EFP} \equiv  \sum_{l=1}^m 
\delta {\cal H}^{(l)} ~+~ V^{LDA}_{xc}$, are independent of the LDA.
%
%

One may wonder why this approach,
based on an argument for non-interacting systems, would give rapid 
convergence in the interacting case.
To see this, we consider the combined action of the many-body Hamiltonian 
${\cal H}$ and the Jastrow factor ${\cal J}$ on the determinant $D$ by defining
${\cal H}_J D(R) \equiv  [{\cal H} {\cal J}(R) D(R)] /{\cal J}(R)$.
The eigenvalues of ${\cal H}$ and ${\cal H}_J$ are identical 
and the eigenfunctions $\Psi$ of ${\cal H}$ satisfy $\Psi = {\cal J} f$
where $f$ is the corresponding eigenfunction of $H_J$.
With a suitable choice of ${\cal J}$, the two-body terms in ${\cal H}_J$ 
can be made weak~\cite{BP}.
For the uniform electron gas, Bohm and Pines determined the 
long-wavelength part of the two-body Jastrow factor to remove the long-range 
fluctuations of the two-body interaction in ${\cal H}_J$~\cite{BP}.
Similarly, the short-range cusp condition~\cite{cusps} 
removes the $e^2/r_{ij}$ divergence in 
${\cal H}_J$.
If ${\cal H}_J$ were truly a non-interacting Hamiltonian  ${\cal H}_{\rm eff}$,
its exact eigenfunctions would be Slater determinants $D$.  
Thus, the motivation for 
choosing a trial Jastrow-Slater wave function
${\cal J} D$
is directly related to the approximation that two- and higher-body 
terms in ${\cal H}_J$ can be neglected~\cite{BP}.
This, in turn, motivates our iterative approach to the
solution of the Euler-Lagrange equations (\ref{eqn3}).

In order to specify the full numerical implementation of the method,
we need to choose an appropriate set of one-body operators ${\cal O}_k$.
We consider three possible choices:

(1)
${\cal O}_k$ is a local
potential $f_k({\bf r})$ so that ${\cal O}_k(i) =  \sum_{j=1}^N f_k({\bf r}_j)$ 
for the configuration $R(i) = ({\bf r}_1,\dots,{\bf r}_N)$. 
The evaluation of such an operator and of the averages required in 
Eq.~\ref{eqn5}, and the increment of $V_{ext}$ by
$\delta V({\bf r}) = \sum_{k=1}^n V_k f_k({\bf r})$ in the
self-consistent LDA calculation is then straight-forward.
This case corresponds to the variational freedom of multiplying all
orbitals $\phi_i$ in $D$ by a common function $1 + \eta_k f_k({\bf r})$ 
and is equivalent to minimizing the energy with respect to the one-body
term in the Jastrow factor~\cite{F}.
 
(2) ${\cal O}_k$ is an angular-momentum-dependent 
potential $f_k(r) {\cal P}_l$ with ${\cal P}_l$ the projection operator 
for angular momentum $l$. The evaluation of the coefficients in Eq.~\ref{eqn5}
can be made using the standard methods for the integration of non-local 
pseudopotentials in VMC~\cite{FWL1} and incrementing the external potential
in the LDA code by an angular-momentum-dependent potential is 
also straight-forward.
This case corresponds to the variational freedom of multiplying 
different angular momentum orbitals $\phi_i$ in $D$ by different
factors.

(3) Arbitrary  variations of the orbitals $\phi_i$ may be allowed
by using the one-body operators ${\cal O}_{ji} = c^\dagger_jc_i$, where 
$i$ labels an occupied orbital of the determinant $D$ and $j$ an 
unoccupied orbital.
${\cal O}_{ji}$ acting on $D$ simply replaces the orbital
$\phi_i$ in $D$ with the orbital $\phi_j$. The averages in 
Eq.~\ref{eqn5} can be efficiently calculated using the relations in
Ref.~\cite{CCK} for replacing a row in a Slater determinant.
The Hamiltonian in the self-consistent LDA calculation is then
incremented by 
$\delta {\cal H}=\sum_{ji} V_{ji} |\phi_j\rangle\langle\phi_i|+c.c.$. 

Thus, while the approach can use, with trivial modifications, all the 
computational techniques to determine LDA orbitals, the final orbitals 
$\phi^{(m)}_i$ minimize the energy for the many-body wave function 
${\cal J} D^{(m)}$ with no restriction on the form of ${\cal J}$.
%
%

\noindent
\begin{minipage}{3.375in}
\begin{figure}
\centerline{\epsfxsize=8.0 cm \epsfysize=6.5 cm \epsfbox{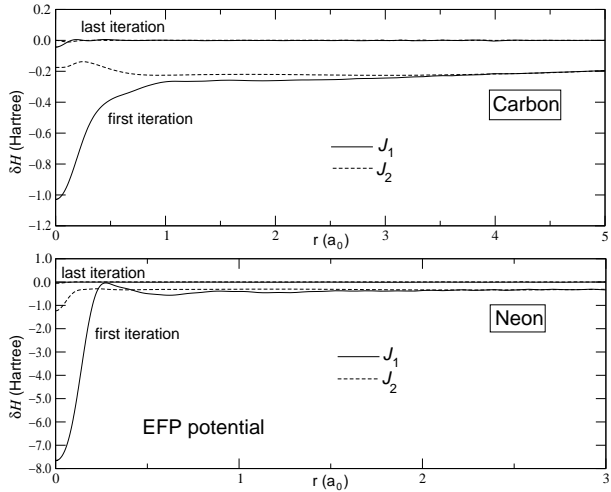}}
\vspace{.2cm}
\caption[]{The incremental EFP potentials, $\delta {\cal H}(r)$,
for the ground state of the carbon and neon pseudo-atoms. 
The potentials corresponding to the Jastrow factors ${\cal J}_1$ and 
${\cal J}_2$ are shown for the first and last iteration.
$\delta {\cal H}(r)$ at the first iteration is calculated using LDA 
orbitals in the determinant.}
\vspace{.2cm}
\label{fig1}
\end{figure}
\end{minipage}
In this Letter, we present results for the
carbon and neon pseudo-atoms (pseudopotentials are used to 
eliminate the $1s$ core electrons~\cite{FWL1}).
The application of the approach to extended systems will be discussed in 
detail elsewhere~\cite{FF}.
In optimizing the orbitals, we have investigated method (1)
with a local EFP potential,
and method (2) with both s and p EFP potentials.
Because there is only one type of s orbital (spin up and down) and only
one radial function for the p orbitals, methods (2) and (3) are
equivalent and all variations of the orbitals 
consistent with the ground state symmetry are allowed in method (2).
For simplicity, we will focus on results obtained using the 
local potentials (1). Interestingly,
the orbitals, charge density, and energies differ only 
very slightly when the full variational freedom of the orbitals 
is allowed using separate s and p non-local potentials.

The initial atomic orbitals in the Jastrow-Slater wave function are 
determined from a LDA calculation.
Because of self-interaction in the LDA, $\delta {\cal H}({\bf r})$ computed 
at the first iteration must behave like $-e^2/r$ at large distances.
Since, at large radii, the sampling of $\delta {\cal H}$ has large
statistical noise due to the very low electron density,
we constrain $\delta {\cal H}^{(1)}(r)$ to behave like $-e^2/r$ at
large $r$ while allowing full variational freedom at smaller $r$.
This is achieved by writing 
$\delta {\cal H}^{(1)}(r) = V_0(r) + \sum_{k=1}^{n_f} V_k \, f_k(r)$,
where $f_k(r)=\cos[(k-1) \pi r/r_c]\exp[-(r/r_c)^4]$
and $V_0(r)$ goes like $-e^2/r$ for $r>r_c$ and smoothly becomes
constant for $r<r_c$.
The parameters $V_k$ are determined by least-square
fitting, as in Eq.~\ref{eqn4}.  After the first iteration,
$V_0(r)$ is not included in fitting $\delta {\cal H}^{(l)}$.

We performed the calculations for two different types of Jastrow 
factor, ${\cal J}_1$ and ${\cal J}_2$.
The Jastrow factor ${\cal J}_1$ only contains electron-nucleus and
electron-electron terms (see Appendix A of Ref.~\cite{MFB}) and
the value of its single free parameter was determined by minimizing 
the energy.  ${\cal J}_2$ includes electron-electron,
electron-electron-nucleus and electron-nucleus terms (modified from
Ref.~\cite{FU} to deal with a pseudo-atom) and variance minimization 
was used to optimize its 25 parameters.

In Fig.~\ref{fig1}, we show the first and last incremental EFP potentials
for carbon and neon, obtained using the two Jastrow factors ${\cal J}_1$ 
and ${\cal J}_2$.
The cut-off radius $r_c$ is equal to 3 a.u.\ for neon and 5 a.u.\ 
for carbon and the number of basis functions, $n_f$, is always less than
50. In each case, the final iteration is almost indistinguishable from zero, 
except for statistical sampling noise near the origin. The potential depends on 
the choice of the correlated component of the wave function: the superiority 
of ${\cal J}_2$ over ${\cal J}_1$ is reflected in the much smaller initial 
potential $\delta {\cal H}^{(1)}(r)$.
\noindent
\begin{minipage}{3.375in}
\begin{table}[htb]
\caption[]{Total energies in VMC (E$_{\rm VMC}$) and DMC (E$_{\rm DMC}$) 
and root mean square fluctuation ($\sigma$) of the local energy in VMC.
$E^{\rm V}_c$ and $E^{\rm D}_c$ are the percentages of correlation
energies recovered in VMC and DMC.  Hartree units are used.
(For carbon, the HF energy is $E_{\rm HF}=-5.3530$ and the ground state 
energy~\cite{dolg} is $E_0=-5.4561$ Hartree. For neon, $E_{\rm HF}=-34.6930$ 
and $E_0 = -35.0106$ Hartree~\cite{dolg}). 
}
\label{tab1}
\begin{tabular}{lcccccc}
 & $\Psi$ & $E_{\rm VMC}$ & $E_{\rm DMC}$ & $E^{\rm V}_c$(\%) & 
$E^{\rm D}_c$(\%) & $\sigma$ \\[.1cm]
\hline\\[-.2cm]
C  & ${\cal J}_1D_{\rm LDA}$ & -5.4345(1)  & --          & 79.1(1) & -- & 0.255 \\
   & ${\cal J}_1D_{\rm EFP}$ & -5.4376(1)  & --          & 82.0(1) & -- & 0.249 \\[.1cm]
   & ${\cal J}_2D_{\rm LDA}$ & -5.4371(1)  & -5.4451(1)  & 81.7(1) & 89.3(1) & 0.219 \\
   & ${\cal J}_2D_{\rm EFP}$ & -5.4373(1)  & -5.4451(1)  & 81.8(1) & 89.3(1) & 0.215 \\[.2cm]
Ne & ${\cal J}_1D_{\rm LDA}$ & -34.9554(2) & --          & 82.6(1) & -- & 1.150 \\
   & ${\cal J}_1D_{\rm EFP}$ & -34.9674(3) & --          & 86.4(1) & -- & 0.882 \\[.1cm]
   & ${\cal J}_2D_{\rm LDA}$ & -34.9912(2) & -35.0041(2) & 93.9(1)& 98.0(1) & 0.630 \\
   & ${\cal J}_2D_{\rm EFP}$ & -34.9913(2) & -35.0040(2) & 93.9(1)& 97.9(1) & 0.624 \\
\end{tabular}
\end{table}
\end{minipage}
From Table~\ref{tab1}, 
we see that for each atom and for each type of Jastrow factor, the
energy is lowered in going from LDA to EFP orbitals.
Since ${\cal J}_2$ has greater 
flexibility than ${\cal J}_1$, this lowering of energy 
is negligible for ${\cal J}_2$. 
For carbon (an open shell system) a multideterminant wave function
is required to accurately represent the correlations and the use 
of a more flexible Jastrow factor ${\cal J}_2$ gains little
over ${\cal J}_1$.
(The very small difference in energy between 
${\cal J}_2 D_{\rm EFP}$ and ${\cal J}_1 D_{\rm EFP}$ may be due either to 
the different parametric forms of ${\cal J}_1$ and ${\cal J}_2$ or
to intrinsic differences between variance minimization 
and energy minimization.) 
In DMC, the energy gain in using EFP instead of LDA orbitals is 
negligible for both systems.
\noindent
\begin{minipage}{3.375in}
\begin{figure}[hbt]
\noindent
\centerline{\epsfxsize=8.0 cm \epsfysize=6.5 cm \epsfbox{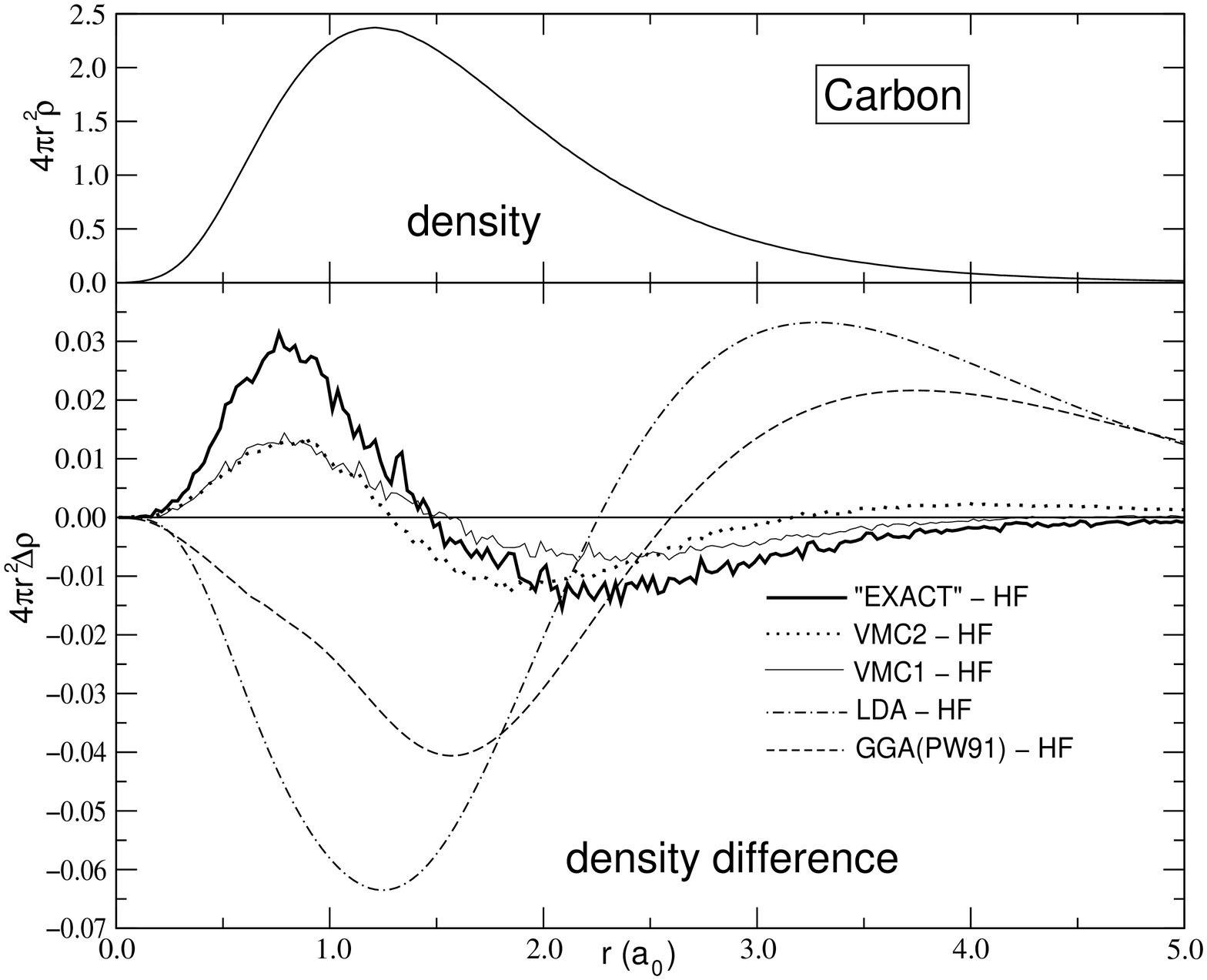}}
\vspace*{.2cm}
\centerline{\epsfxsize=8.0 cm \epsfysize=6.5 cm \epsfbox{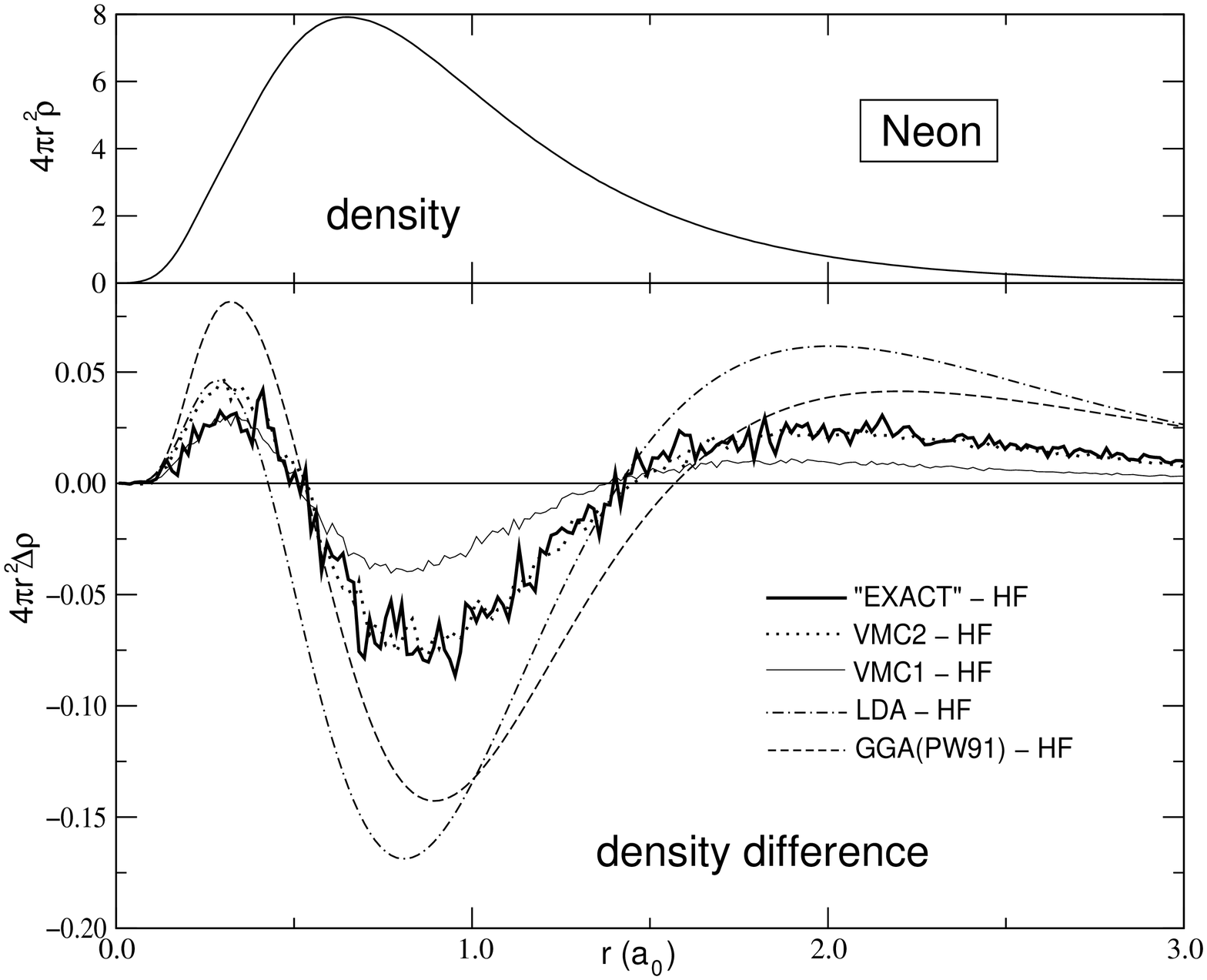}}
\caption[]{Valence charge density $4\pi r^2\rho(r)$ for carbon and neon. 
The upper panel for each atom shows the ``exact'' density\cite{dmcrho}.
The lower panel shows the differences from the HF density of the 
``exact'', LDA, GGA (PW91) and the densities for the wave functions 
${\cal J}_1D_{\rm EFP}$ (VMC1) and ${\cal J}_2D_{\rm EFP}$ (VMC2).}
\vspace{.2cm}
\label{fig2}
\end{figure}
\end{minipage}
In Fig.~\ref{fig2}, the densities are shown for carbon and neon. 
In both systems and for both Jastrow factors, the 
VMC density obtained with the EFP approach is substantially closer to the best 
estimate of the true density than HF, and much closer than either LDA or 
the Perdew-Wang '91 (PW91) generalized gradient approximation (GGA)~\cite{gga}.  
In carbon, neither ${\cal J}_1$ nor ${\cal J}_2$ can capture 
the intrinsic multi-configuration correlation and the accuracy in
the densities cannot rival that obtained in neon with ${\cal J}_2D_{\rm EFP}$.

In conclusion, we have demonstrated, for the first time, a numerically 
stable, rapidly convergent method which combines Monte Carlo sampling with 
existing self-consistent field techniques to minimize the 
energy with respect to the orbitals in 
a correlated Jastrow-Slater wave function.
The approach may be combined with variance minimization methods for the 
optimization of the Jastrow factor.
The resulting variational many-body wave functions 
have electron densities very close to
the most accurate densities available for atoms, using variance
minimization and DMC methods.
Preliminary tests show that the approach, in modified form, is also
applicable to multi-determinant wave functions.
We thank C. Umrigar for useful discussions and E. Shirley for the
use of his Hartree-Fock code.
This work was supported by Enterprise Ireland, Contract SC/98/748.

\end{multicols}


\begin{references}

\bibitem{CA} D. M. Ceperley and B. J. Alder, Phys. Rev. B {\bf 36}, 2092 (1987).
\bibitem{UWW} C. J. Umrigar, K. G. Wilson, and J. W. Wilkins, Phys. Rev. Lett. {\bf 60}, 1719 (1988).
\bibitem{FWL1} S. Fahy, X. W. Wang, and S. G. Louie, Phys. Rev. B {\bf 42}, 3503 (1990).
\bibitem{LCM}  X. P. Li, D. M. Ceperley, and R. M. Martin, Phys. Rev. B {\bf 44}, 10929 (1991).
\bibitem{CMR} J. C. Grossman, L. Mitas, and K. Raghavachari, Phys. Rev. Lett. {\bf 75}, 3870 (1995).
\bibitem{R} G. Rajagopal {\it et al.},  Phys. Rev. B {\bf 51}, 17698 (1995).
\bibitem{FU} C. Filippi and C. J. Umrigar, J. Chem. Phys. {\bf 105}, 213 (1996).
\bibitem{W} A. J. Williamson {\it et al.},  Phys. Rev. B {\bf 53}, 9640 (1996).
\bibitem{H} A. Harju {\it et al.}, Phys. Rev. Lett. {\bf 79}, 1173 (1997).
\bibitem{EuLa} Approximate solutions have been given for certain systems
within the Fermi hypernetted-chain approach. See, for example,
E. Krotschek, W. Kohn, and G.-X. Qian, Phys. Rev. B {\bf 32}, 5693 (1985).
\bibitem{BP} D. Bohm and D. Pines, Phys. Rev. {\bf 92}, 609 (1953).
\bibitem{F} S. Fahy, in {\it Quantum Monte Carlo Methods in Physics and
Chemistry}, pgs. 101-27, eds. M. P. Nightingale and C. J. Umrigar,
NATO Science Series, Vol. C 525 (Kluwer, Dordrecht, 1999).
\bibitem{CCK} D. Ceperley, G. V. Chester, and M. H. Kalos, Phys. Rev. B {\bf 16}, 3081 (1977).
\bibitem{FF} S. Fahy and C. Filippi, to be published.
\bibitem{MFB} A. Malatesta, S. Fahy, and G. B. Bachelet, Phys. Rev. B {\bf 56}, 12201 (1997).
\bibitem{PZ} J. P. Perdew and A. Zunger, Phys. Rev. B {\bf 23}, 5048 (1981).
\bibitem{gga} J. P. Perdew in {\it Electronic structure of solids '91},
edited by P. Ziesche and H. Eschrig (Akademie Verlag, Berlin, 1991).
\bibitem{cusps} T. Kato, Commun. Pure Appl. Math {\bf 10}, 151 (1957).
\bibitem{dolg} M. Dolg, Chem. Phys. Lett. {\bf 250}, 75 (1996).
The cited valence correlations energies of 0.1031 for carbon and 0.3176 Hartree
for neon are added to the HF energy to obtain an estimate of the ground state
energy.
\bibitem{dmcrho} The ``exact'' density is the second order estimator of the 
density using, for carbon, a multi-configuration wave function which includes 
$d$-excitations optimized by variance minimization ($E_{\rm VMC}=-5.45061(2)$, 
$E_{\rm DMC}=-5.4544(1)$, $\sigma=0.154$ Hartree) and, for neon, 
the wave function ${\cal J}_2D_{\rm EFP}$.

\end{references}
\end{document}